\documentclass[usenatbib,useAMS,apj,iop,tighten]{emulateapj}
\usepackage{epsfig,aas_macros}
\usepackage{amsmath,amssymb,bm}
\usepackage{color}
\usepackage{aas_macros}
\usepackage[varg]{txfonts}

\begin{document}
\label{firstpage}

\title[On the Formation of Super-Earths]{On the Formation of
  Super-Earths with Implications for the Solar System}

\author{Rebecca G. Martin} 
\author{Mario Livio}
\affil{Department of Physics and Astronomy,
  University of Nevada, Las Vegas, 4505 South Maryland Parkway, Las
  Vegas, NV 89154, USA }

\begin{abstract}
We first consider how the level of turbulence in a protoplanetary disk
affects the formation locations for the observed close--in
super--Earths in exosolar systems. We find that a protoplanetary disk
that includes a dead zone (a region of low turbulence) has
substantially more material in the inner parts of the disk, possibly
allowing for {\it in situ }formation. For the dead zone to last the
entire lifetime of the disk requires the active layer surface density
to be sufficiently small, $\Sigma_{\rm crit}\lesssim 100\,\rm
g\,cm^{-2}$. Migration through a dead zone may be very slow and thus
super--Earth formation followed by migration towards the star through
the dead zone is less likely. For fully turbulent disks, there is not
enough material for {\it in situ} formation. However, in this case,
super--Earths can form farther out in the disk and migrate inwards on
a reasonable timescale. We suggest that both of these formation
mechanisms operate in different planetary systems. This can help to
explain the observed large range in densities of super--Earths because
the formation location determines the composition.  Furthermore, we
speculate that super--Earths could have formed in the inner parts of
our solar system and cleared the material in the region inside of
Mercury's orbit. The super--Earths could migrate through the gas disk
and fall into the Sun if the disk was sufficiently cool during the
final gas disk accretion process. While it is definitely possible to
meet all of these requirements, we don't expect them to occur in all
systems, which may explain why the solar system is somewhat special in
its lack of super--Earths.
\end{abstract}

\keywords{accretion, accretion disks -- protoplanetary disks --
  planet--disk interactions -- planetary systems: formation}

\section{Introduction}
\label{intro}

Recently we identified the two most unusual aspects of our solar
system (compared to all the observed exoplanetary systems) to be,
first, the lack of super-Earths \citep[planets with a mass in the
  range $1-10\,\rm M_\oplus$, e.g.][]{Valencia2007} and, secondly, the
lack of planets inside of Mercury's orbit \citep{Martin2015}.  More
than half of the observed Sun-like stars in the solar neighborhood
have one or more super--Earth planets on  low eccentricity orbits with
periods of days to months \citep{Mayor2011,Batalha2013,
  Fressin2013,Burke2015}. This is in contrast to our solar system
which is depleted in mass in this region. To examine potential reasons
for this difference, in this work we consider formation processes for
super--Earths and investigate the conditions in the solar nebula that
could have affected the outcome for our solar system.

The timeline for the formation of the planets in our solar system is
thought to have been as follows. The planetesimals formed within a few
million years of the birth of the Sun \citep{Connelly2012}. The giant
planets formed quickly allowing for the accretion of material from the
gas disk \citep[e.g.][]{Alibert2005}. However, the terrestrial planets
formed long after the gas disk was dispersed, on a timescale of around
$10-100$ million years \citep[e.g.][]{Kenyon2006}. There are no
super--Earths in our solar system so understanding their formation on
the basis of solar system data alone is more difficult.

\begin{figure}
\begin{center}
\includegraphics[width=8.4cm]{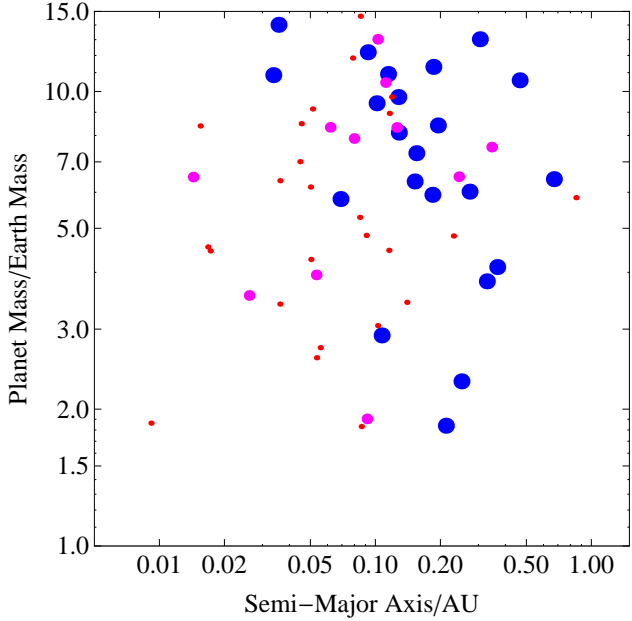}
\end{center}
\caption{Planet mass and semi--major axis of the observed exoplanets
  with a mass in the range $1\,{\rm M_\oplus}<M_{\rm p}<15\,\rm
  M_\oplus$ that have a density measurement. The large (blue) points
  denote planets with a low density $\rho \le 1.6\,\rm g\,cm^{-3}$,
  the medium--size (magenta) points denote those with a density in the
  range $1.6{\,\rm g\,cm^{-3}}<\rho<3.9\,\rm g\,cm^{-3}$ and the small
  (red) points denote those with a high density, $\rho>3.9\,\rm
  g\,cm^{-3}$. For reference, in our solar system the terrestrial
  planets have average densities $\rho>3.9\,\rm g\,cm^{-3}$ and the
  giant planets have average densities $\rho<1.6\,\rm
  g\,cm^{-3}$. Data are from exoplanets.org.}
\label{density}
\end{figure}

The observed close--in super--Earths exhibit a wide range of densities
\citep[e.g.][]{Wolfgang2012, Howe2014,Knutson2014,Marcy2014}
suggesting that there may be several different mechanisms for their
formation \citep[see also Figure~5 in][]{Martin2015}. In
Fig.~\ref{density} we show the planet mass and semi--major axis of
observed exoplanets that have a density measurement \citep[the data
  are taken from exoplanets.org,][]{Han2014}. The largest points show
the low density planets (planets with densities similar to the giant
planets in our solar system, density $\rho<1.6\,\rm g\,cm^{-3}$), the
medium--size points have density in the range $1.6{\,\rm
  g\,cm^{-3}}<\rho<3.9\,\rm g\,cm^{-3}$ and the small points show the
high density planets (with density greater than $3.9\,\rm g\,cm^{-3}$,
similar to the terrestrial planets in our solar system). There appears
to be little correlation between the density and the planet
mass.\footnote{There is much discussion on this point in the
  literature \citep[see
    e.g.][]{Petigura2013,Lopez2014,Marcy2014b,Weiss2014,Morton2014}.}
There is a slight correlation between the density and the semi--major
axis in that there are fewer low density planets close to the
star. The compositions of extrasolar super-Earths suggest that at
least some of them have substantial gaseous atmospheres. This is at
odds with the timescale for terrestrial planet formation and therefore
(at least some) super-Earths likely formed while the gas disk was
still present.  The composition of the planets is dependent on their
formation location. Planets that form outside of the water snow line
\citep[the radial location from the star where the temperature is low
  enough for water to become solid,
  e.g.][]{Lecaretal2006,MartinandLivio2012,MartinandLivio2013snow}
will be water--rich while those that form close to their star will
likely be more rocky. We expect that planets that form farther out in
the disk may be less dense than those that form close to the
star. However, accretion of gaseous material from the disk may
significantly reduce the final average density of the planet and thus
the time of planet formation is also important.
%Therefore, determining the composition of super--Earths
%will help to constrain their formation site.

There are two general orbital locations suggested for the formation of
super-Earths. Either they formed {\it in situ} with no significant
migration, or else they formed similarly to giant planets, outside the
snow line in the protoplanetary disk and then migrated inwards. Giant
planets with short orbital periods are thought to have formed farther
out and migrated inwards to their current locations
\citep[e.g.][]{Bodenheimer2000}. However, hot Jupiters are only found
around about 1\% of stars
\citep[e.g.][]{Mayor2011,Howard2012,Wright2012, Batalha2013}.
Typically their orbits are highly misaligned, suggesting that they
were driven to these locations through secular perturbations or
planet--planet scattering \citep[e.g.][]{Takeda2005, FordRasio2008,
  Nagasawa2008, Perets2009}. On the other hand, super--Earths could
have an occurrence rate as high as 50\% or more
\citep{Fressin2013}. Thus, the hot Jupiters probably migrated to their
observed location while super--Earths, that are far more abundant, may
be more likely to have formed {\it in situ}.  We discuss both
possibilities here.

%There appear to be two possible mechanisms for super--Earth formation.
%The host stars for the lowest period planets are more metal rich than
%those with longer orbital periods \citep{Zhu2015}. In
%\cite{Martinetal2012a} we found that the extend of the dead zone is
%not affected by the metallicity, assuming that Ohmic diffusion is most
%important. However, the higher metallicity implies there are more
%solids available for planet formation.

%\section{Super--Earth Formation Locations}

%We consider here the possible formation locations for super--Earths
%within the protoplanetary gas disc and how those fit in with
%observed properties.

%\subsection{{\it In situ} Formation of Super--Earths}

\cite{Chiang2013} constructed a minimum-mass extrasolar nebula (MMEN)
from observations of super--Earth exoplanets with orbital periods
$P<100\,\rm$days. This is a circumstellar disk of solar composition
that allows {\it in situ} formation of close--in super--Earths. They
found a surface density for the gas disk of
\begin{equation}
\Sigma_{\rm MMEN}=9900\,\left(\frac{R}{1\,\rm AU}\right)^{-1.6}\,\rm g\,cm^{-2},
\end{equation}
where $R$ is the radial distance from the central star.  This is somewhat
higher than the minimum mass solar nebular (MMSN), required to form
the planets in our solar system, which is given by
\begin{equation}
\Sigma_{\rm MMSN} = 1700\,\left(\frac{R}{1\,\rm AU}\right)^{-1.5}\,\rm g\,cm^{-2}
\end{equation}
\citep{Weidenschilling1977, Hayashi1981}. However, the MMSN is not
necessarily applicable inside of Mercury's orbit at $0.4\,\rm AU$.
%so it is not clear
%whether such a comparison can really be made.

\cite{Hansen2012} used n--body simulations to form super-Earth planets
     {\it in situ}. They found it was possible if the amount of rocky
     material interior to $1\,\rm AU$ is about $50-100\,\rm
     M_\oplus$. They suggested that this would require significant
     radial migration of solid material before the end stages of
     planet formation.  The minimum--mass solar nebular
     \citep{Weidenschilling1977, Hayashi1981} has only $3.3\,\rm
     M_\oplus$ in solids interior to $1\,\rm AU$, although the planet
     formation process is not likely to be completely efficient.  The
     ratio of dust to gas is often quoted to be 0.01 in the
     ISM. However, in accretion disks observations indicate that it
     may be much higher \citep[e.g.][]{Williams2014}. The mass in the
     gas disk in this region would need to be in the range
     $0.0015-0.03\,\rm M_\odot$.  An advantage of the {\it in situ}
     formation model is that it can explain several properties of the
     observed planet inclination and eccentricity distributions as
     well as the orbital spacing \citep{Hansen2012,Hansen2013}.

%\subsection{Migration of Super--Earths}

The alternative super--Earth formation theory suggests that they form
farther out in the disk where there is more solid material and then
migrate inwards \citep{Terquem2007,McNeil2009,Ida2010,Cossou2014}.  In
this scenario, planets grow through Earth size embryo--embryo
collisions. Smooth migration means that the planets form in chains of
mean motion resonances.  However, multiplanet systems observed with
Kepler generally do not show planets locked into resonances
\citep[e.g.][]{Fabrycky2012,Batygin2013,Steffen2015} and thus the
resonances must be broken after planet formation
\citep[e.g.][]{Rein2012,Goldreich2014}. \cite{Cossou2014} suggested that
hot super--Earths and giant planet cores form in the same way. They
both migrate inwards but the giant planet cores become massive enough
for the direction of the migration to reverse. The super--Earths pile
up at the inner edge of the disk.

In this work we present a gas disk model that allows for the formation
via both mechanisms in both locations, depending upon the disk
properties.  In Sections~\ref{turb} and~\ref{dead} we construct
numerical models of protoplanetary disks without and with a dead zone
(a region of low turbulence). This will allow us to draw some
conclusions about the formation mechanisms for super--Earths. In
Section~\ref{solarsystem} we discuss the implications for our own
solar system and we discuss and summarize our results in
Sections~\ref{discussion} and~\ref{conc}.

%\section{Gas disk Models}
%\label{models}

%In this Section we consider the evolution of the protoplanetary gas
%disk to investigate the effect of a dead zone on the amount of
%material in the super--Earth formation region. We first consider fully
%turbulent disk models in Section~\ref{turb} and then we consider a
%range of dead zone parameters in Section~\ref{dead}.

\section{Fully Turbulent disk Model}
\label{turb}

In this Section we consider the evolution of a fully turbulent
protoplanetary gas disk.
%%%%%%%%%%%%%%%%%%%%%%%%%%%%%%%%%%%%%%%%%%%%%%%%%%%%%
The disk is in Keplerian rotation with angular velocity
$\Omega=\sqrt{G M/R^3}$ around a central mass $M$ at radial distance
$R$. Turbulence, driven by the magneto--rotational instability (MRI),
drives a kinematic turbulent viscosity
\begin{equation}
\nu=\alpha\frac{c_{\rm s}^2}{\Omega},
\label{nua}
\end{equation}
where $\alpha$ is the \cite{SS1973} viscosity parameter, the sound
speed is $c_{\rm s}=\sqrt{{\cal R} T_{\rm c}/\mu} $ with mid--plane
temperature $T_{\rm c}$,  ${\cal R}=8.31\times 10^7\,\rm erg\,K^{-1}\rm
mol^{-1}$ is the gas constant and $\mu=2.3\,\rm g\,mol^{-1}$ is the
mean molecular weight. The evolution of the surface density, $\Sigma$,
is governed by
\begin{equation}
\frac{\partial \Sigma}{\partial t}=
\frac{3}{R}\frac{\partial}{\partial  R}
\left[ R^\frac{1}{2}\frac{\partial}{\partial R}\left( \xi R^\frac{1}{2}\right)\right],
\label{sigma}
\end{equation}
\citep[e.g.][]{LP1974, Pringle1981} where $\xi=\nu \Sigma$ in this
fully MRI turbulent disk.

The temperature evolves according to the simplified energy equation
\begin{equation}
\frac{\partial T_{\rm c}}{\partial t}=\frac{2(Q_+-Q_-)}{c_{\rm p}
  \Sigma}
\label{temp}
\end{equation}
\citep{Pringleetal1986,Cannizzo1993}.  The disk specific heat for
temperatures around $10^3\,\rm K$ is $c_{\rm p}=2.7 {\cal R}/\mu$.
The disk is heated by viscous dissipation according to
\begin{equation}
Q_+=\frac{9}{8}\Omega^2 \xi.
 \label{Qp}
\end{equation}
We assume that each annulus of the disk radiates as a black body and
so the local cooling is
\begin{equation}
Q_-=\sigma T_{\rm e}^4,
\end{equation}
where $T_{\rm e}$ is the temperature at the surface of the disk and
$\sigma$ is the Stefan-Boltzmann constant.  Assuming energy balance, the mid--plane temperature
and the surface temperature are related through
\begin{equation}
T_{\rm c}^4=\frac{3}{4}\tau T_{\rm e}^4,
\end{equation}
where the optical depth is
\begin{equation}
\tau=\kappa(T_{\rm c}) \frac{\Sigma}{2}.
\end{equation}
We use the simplified opacity of \cite{Armitage2001}
$\kappa(T)=0.02\,T^{0.8} \,\rm cm^2/g$ that is valid in the inner
parts of the disk.

%%%%%%%%%%%%%%%%%%%%%%%%%%%%%%%%%%%%%%%%%%%%%%%%%%%%% 

At early times in the protostellar accretion history the infall
accretion rate on to the disk is expected to be around $\dot M_{\rm
  infall} \approx c_{\rm s}^3/G$, \citep[e.g.][]{Larson1969, Shu1977,
  Basu1998}. For temperatures around $10\,\rm K$, the infall accretion
rate is of the order of $10^{-5}\,\rm M_\odot\,\rm yr^{-1}$. We first
run a disk model with a constant infall accretion rate of $1\times
10^{-5}\,\rm M_\odot\,yr^{-1}$ until it reaches a steady state. This
is the initial disk setup that we use for all of the simulations.  The
infall accretion rate on to the disk evolves with time as
\begin{equation}
\dot M_{\rm infall}=\dot M_{\rm i}\exp \left(-\frac{t}{t_{\rm ff}}\right),
\label{mdot}
\end{equation}
where the initial accretion rate is $10^{-5}\,\rm M_\odot\,yr^{-1}$
and $t_{\rm ff}=10^5\,\rm yr$
\citep[see][]{Armitage2001,Martinetal2012b}.

\subsection{{\it In Situ} Super--Earth Formation}

\begin{figure}
\begin{center}
\includegraphics[width=7.0cm]{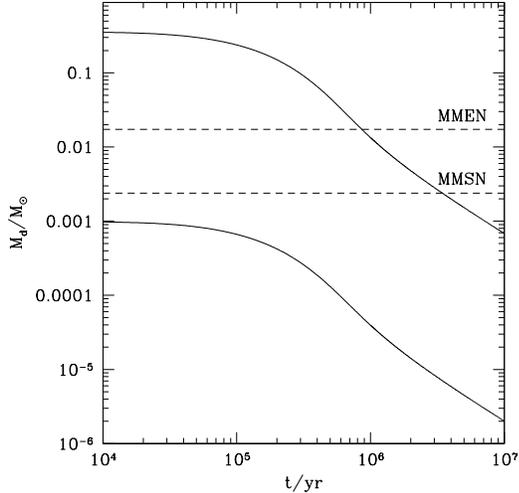}
\end{center}
\caption{Disk mass up to a radius of $R=1\,\rm AU$ as a function of
  time (lower line) and total disk mass up to a radius of $40\,\rm AU$
  (upper line). The initial disk surface density is a steady state
  fully turbulent disk with an infall accretion rate of $1\times
  10^{-5}\,\rm M_\odot\,yr^{-1}$. The infall accretion rate decreases
  exponentially in time according to equation~(\ref{mdot}). The dashed
  lines show the mass in $R<1\,\rm AU$ for the MMSN (lower) and the
  MMEN (upper). }
\label{fig1}
\end{figure}

We integrate the disk evolution equations~(\ref{sigma})
and~(\ref{temp}) for a disk that extends from $R_{\rm in} = 2.33
\times 10^{−3}\,\rm AU$ to $R_{\rm out} = 40\,\rm AU$ around a solar
mass star. The grid contains $200$ points distributed uniformly in
$R^\frac{1}{2}$. Material is added to the disk at a radius of $35\,\rm
AU$. At the inner edge of the disk a zero torque boundary condition
allows the inward flow of gas out of the grid and toward the central
star. The flow is prevented from leaving the outer boundary with a
zero radial velocity outer boundary condition.

We consider a fully turbulent disk model that has a viscosity
parameter $\alpha=0.01$ everywhere in the disk
\citep[e.g.][]{Hartmann1998}. Figure~\ref{fig1} shows the mass of the
disk inside of $R=1\,\rm AU$ (lower line) and the total disk mass
  up to $R=40\,\rm AU$ (upper line) as a function of time.  The
material in $R<1\,\rm AU$ is potentially available for the formation
of super-Earth planets in these inner regions. Note that the amount of
mass in the MMSN in this region is $0.002\,\rm M_\odot$. It seems a
little surprising that the fully turbulent disk model is never above
this value. This is partially because the MMSN is not applicable
inside of Mercury's orbit. The MMSN has a very steep dependence on
radius and thus predicts a rather high mass in the inner regions of
the disk. Similarly, the MMEN also predicts a very high mass in the
inner regions in order to be able to form the observed exoplanets. We
can conclude that the fully turbulent disk model probably cannot form
the planets that are observed close to their host stars {\it in situ}.
% We discuss this
%possibility further in Section~\ref{discussion}.

\subsection{Migration of Super--Earths}

While {\it in situ} formation of super--Earths seems to be ruled out
in the case of a fully turbulent disk, formation of super--Earths
farther out in the disk followed by inward migration remains a
possibility in the fully turbulent disk model. There are two types of
planetary migrations depending on the mass of the planet and the
properties of the disk. In type~I migration the planet is not massive
enough to open a gap in the disk and the surface density of the disk
remains largely unperturbed by the presence of the planet. This type
of migration is not dependent on the viscosity of the disk.  In a
fully turbulent disk, super--Earths are not large enough to open a gap
in the disk and so they migrate via type~I migration.  The timescale
for this migration is given approximately by
\begin{align}
\tau_{\rm type~I}\approx & \,\,\,6.7\times 10^5 \left(\frac{M_{\rm p}}{5\,\rm M_\oplus}\right)^{-1} \left(\frac{M}{1\,\rm M_\odot}\right)^{-\frac{3}{2}} \left(\frac{a}{1\,\rm AU}\right)^{-\frac{1}{2}} \cr
& \times  \left(\frac{\Sigma}{100\,\rm g\rm cm^{-2}}\right)^{-1}\left(\frac{H/R}{0.05}\right)^2 \,\rm yr
\label{type1}
\end{align}
\citep{Tanaka2002,Armitage}, where $M_{\rm p}$ is the mass of the
planet, $a$ is the orbital radius of the planet, $M$ is the mass of
the star and $H/R$ is the disk aspect ratio.  We do note that this
timescale is very sensitive to disk parameters and the direction of
type~I migration is determined by the mass of the planet.

Equation~(\ref{type1}) shows that the inward migration timescale is
around a few $10^5\,\rm yr$ and so the migration can comfortably take
place within the lifetime of the disk. The difficulty is rather how to
stop the migration \citep[e.g.][]{Ida2008}. There has been much
discussion about planet traps that change the direction of type~I
migration
\citep[e.g.][]{Masset2006,Morbidelli2008,Hasegawa2011,Hasegawa2013}. These
can occur at transitions such as snow lines, dead zone boundaries and
heat transitions. The radial location of such a trap may move slowly
in time, thus transporting the planet on a much longer
timescale. Thus, formation farther out followed by inward migration is
possible in principle in the fully turbulent disk model.

\section{Disk Model with a Dead Zone}
\label{dead}

It is generally accepted that protoplanetary disks are not fully
turbulent, but rather that they contain a region of low turbulence, a
dead zone \citep[e.g.][]{Gammie1998,Turner2008,
  Bai2011,Simonetal2011,Okuzumi2011,Dzyurkevich2013}. We therefore
solve the accretion disk equations that include a dead zone region. We
define $\Sigma_{\rm m}$ to be the surface density in the MRI active
layers and
\begin{equation}
\Sigma_{\rm g}= \Sigma-\Sigma_{\rm m}
\label{sigma_g}
\end{equation} 
is the surface density in the dead zone layer. If there is no dead
zone at a particular radius, then $\Sigma_{\rm g}=0$ and $\Sigma_{\rm
  m}=\Sigma$ there. We consider two prescriptions for determining the
surface density of the dead zone:
\begin{enumerate}
\item We assume that the disk surface layers are ionized by external
  sources \citep[cosmic rays or X-rays from the central star,
    e.g.][]{Sano2000,Matsumura2003,Glassgoldetal2004} to a maximum
  surface density depth of $\Sigma_{\rm crit}/2$ on the upper and
  lower disk surfaces, where $\Sigma_{\rm crit}$ is constant
  \citep[e.g.][]{Armitage2001,MartinandLubow2011,Zhuetal2010b}. Cosmic
  rays are thought to ionize about $200\,\rm g\,cm^{-2}$ while X-rays
  are about an order of magnitude smaller.  The disk surface layers
  always contain MRI turbulence.  Furthermore, if the temperature is
  greater than a critical value, $T_{\rm crit}$, then the disk is
  thermally ionized allowing the MRI to operate throughout the
  vertical extent. The value of $T_{\rm crit}$ is not well determined
  and so we consider two values of $800$ and $1400\,\rm K$
  \citep[e.g.][]{Armitage2001,Zhuetal2010a}. Thus, the disk is MRI
  active if either $T_{\rm c}>T_{\rm crit}$ or if $\Sigma<\Sigma_{\rm
    crit}$ and otherwise, there is a dead zone layer at the
  mid--plane.
\item The dead zone surface density is determined via a critical
  magnetic Reynolds number \citep[e.g.][]{Hawley1995,
    Fleming2000,Martinetal2012a,Martinetal2012b}. The disk is ``dead''
  if $Re_{\rm M}<Re_{\rm M,crit}$, where $Re_{\rm
    M}=\sqrt{\alpha}c_{\rm s}H/\eta$, where $H$ is the disk scale
  height and the Ohmic resistivity is $\eta=234\sqrt{T_{\rm c}}/x_{\rm
    e}\,\rm cm^2/s$ \citep{Blaes1994} and $x_{\rm e}$ is the electron
  fraction. We use the analytic approximations for the active layer
  surface density shown in equations~(27) and~(28) in
  \cite{Martinetal2012a} that include thermal ionization, cosmic ray
  ionization and the effects of recombination.
\end{enumerate}

Because the dead zone acts like a plug in the accretion flow, material
accumulates in this region and may become massive enough to become
self--gravitating. Viscosity may be generated by the MRI and by
self--gravity \citep{Paczynski1978,Lodato2004}.  The viscosity driven
by the MRI in the magnetic layers is
\begin{equation}
\nu_{\rm m}=\alpha_{\rm m}\frac{c_{\rm m}^2}{\Omega},
\label{nub}
\end{equation}
where we take $\alpha_{\rm m}=0.01$ and $c_{\rm m}=\sqrt{{\cal R}
  T_{\rm m}/\mu} $ is the sound speed in the magnetic layer with
temperature $T_{\rm m}$. The dead zone is assumed to have zero
turbulence, unless it becomes self--gravitating. However, the
inclusion of a small amount of turbulence in the dead zone
(e.g. $\alpha\lesssim 10^{-3}$) does not affect the behaviour of the
disk significantly \citep{MartinandLubow2013dza}. As material builds
up, the disk becomes self-gravitating if the Toomre parameter
$Q<Q_{\rm crit}=2$, where
\begin{equation}
Q=\frac{c_{\rm g}\Omega}{\pi G \Sigma},
\end{equation} 
and the sound speed at the disk mid-plane is given by $c_{\rm
  g}=\sqrt{ {\cal R}T_{\rm c}/\mu},$. We approximate the temperature
of the self-gravitating region as the mid--plane temperature, $ T_{\rm
  c}$.  Self--gravity drives a viscosity that is approximated by
\begin{equation}
\nu_{\rm g}=\alpha_{\rm g}\frac{c_{\rm g}^2}{\Omega}\left[\left( \frac{Q_{\rm crit}}{Q}\right)^2-1\right]
\label{nug}
\end{equation}
 for $Q<Q_{\rm crit}$ and zero otherwise
 \citep[e.g.][]{Lin1987,Lin1990} and we take $\alpha_{\rm
   g}=\alpha_{\rm m}$.

The surface density and temperature evolution for a disk with a dead
zone is given by equations~(\ref{sigma}) and~(\ref{temp}) but here with
$\xi=\nu_{\rm m} \Sigma_{\rm m} + \nu_{\rm g} \Sigma_{\rm g}$. 
%The
%mid-plane disk temperature for an optically thick disk in thermal
%equilibrium is found by assuming energy balance in a layered model
%\citep{MartinandLubow2011}
%\begin{equation}
%\sigma T_{\rm c}^4 = \frac{9}{8}\Omega^2 \left( \nu_{\rm m} \Sigma_{\rm m}\tau_{\rm m} + \nu_{\rm g} \Sigma_{\rm g} \ta%u \right)
%\label{tau}
%\end{equation}
%and 
The magnetic layer temperature is related to the surface temperature with
\begin{equation} 
T_{\rm m}^4=\frac{3}{4}\tau_{\rm m}T_{\rm e}^4,
\label{rel}
\end{equation} 
where the optical depth to the magnetic region is
\begin{equation}
\tau_{\rm m}=\kappa(T_{\rm m}) \frac{\Sigma_{\rm m}}{2}.
\end{equation}
The optical depth within the dead zone layer is
\begin{equation}
\tau_{\rm g}=\kappa(T_{\rm c})\frac{\Sigma_{\rm g}}{2}
\end{equation}
and
\begin{equation}
\tau = \tau_{\rm m} + \tau_{\rm g}.
\end{equation} 
% Note $\tau_{\rm g}$ is defined even in a dead zone layer ($\nu_{\rm g} = 0$) with $\Sigma_{\rm g}$ defined by equation (\ref{sigma_g}).
%The energy equation~(\ref{temp}) in a steady state has the solution
%\begin{equation}
%\sigma T_{\rm e}^4= \frac{9}{8} \Omega^2 \left( \nu_{\rm m} \Sigma_{\rm m} + \nu_{\rm g} \Sigma_{\rm g} \right).
%\label{heat}
%\end{equation}
%From equations (\ref{tau})-(\ref{heat}), we obtain an expression for the cooling function 
The cooling function is found to be
\begin{equation}
Q_-=\sigma T_{\rm e}^4 = \tau^{-1} \left( \sigma T_{\rm c}^4 + \frac{9}{8} \Omega^2 \nu_{\rm m}\Sigma_{\rm m}  \tau_{\rm g} \right)
\label{Qmg-}
\end{equation}
\citep{MartinandLubow2011}.  Although the disk is not in thermal
equilibrium, we apply this cooling function to equation~(\ref{temp})
and so we do not attempt to treat the cooling during viscosity
transitions consistently.

%%%%%%%%%%%%%%%%%%%%%%%%%%%%%%%%%%%%%%%%%%%%%%%%%%%%%

%We consider two
%different dead zone models. In the first, the disk is non--turbulent
%if the temperature is less than a critical temperature, $T_{\rm
%  crit}$, and the surface density is higher than a critical surface
%density, $\Sigma_{\rm crit}$. Thus, this models assumes that the
%active layer surface density is constant with radius. The second model
%assumes that the disk is ``dead'' if the critical magnetic Reynolds
%number is less than a critical value, $Re_{\rm M}<Re_{\rm M,crit}$.

The initial disk setup is the steady state disk with an infall
accretion rate of $10^{-5}\,\rm M_\odot\,yr^{-1}$. The infall
accretion rate decreases exponentially in time according to
equation~(\ref{mdot}). Thus, the total amount of mass that is accreted
on to the disk is exactly the same in these disk models as in the
fully turbulent disk model described in the previous section.  The
disk with a dead zone is not in a quasi--steady--state as the infall
accretion rate decreases, as the fully turbulent disk model
is. Instead, material building up becomes self--gravitating. The extra
heating by self--gravity can lead to the MRI being triggered within
this region, if the infall accretion rate is sufficiently high. When
this happens, the disk becomes MRI active throughout, leading to a
large amount of material accreting on to the star in a short time
interval \citep{MartinandLubow2013prop}. This is an accretion outburst
that is thought to be the explanation for the observed FU
Orionis--type outbursts
\citep[e.g.][]{Armitage2001,Zhuetal2010b,MartinandLubow2011}. These
outbursts occur during the initial disk evolution but at later times
when the infall accretion rate has dropped there may still be a dead
zone, but there are no further outbursts. Planets that survive must
form after the last outburst otherwise they will likely be swept on to
the central star during the outburst.

\subsection{{\it In Situ} Super--Earth Formation}

\begin{figure}
\begin{center}
\includegraphics[width=8.4cm]{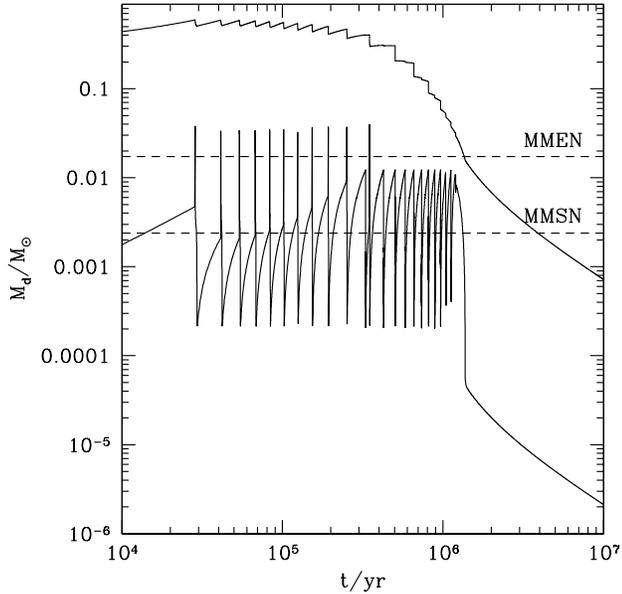}
\end{center}
\caption{Same as Fig.~\ref{fig1} except the disk has a dead zone
  defined by model~1 with $\Sigma_{\rm crit}=200\,\rm g\,cm^{-2}$ and
  $T_{\rm crit}=800\,\rm K$.}
\label{fig2}
\end{figure}

We first consider the possibility of {\it in situ} super--Earth
formation by examining the amount of material in the inner parts of
the disk available for planet formation. We consider a dead zone
defined by model~1 with $T_{\rm crit}=800\,\rm K$ and $\Sigma_{\rm
  crit}=200\,\rm g\,cm^{-2}$.  Fig.~\ref{fig2} shows the evolution of
the amount of material at $R<1\,\rm AU$.  The sharp increases in the
amount of material are when an outburst is triggered and a large
amount of material flows through the region. We suggest that if
super--Earths are to form and survive within such a disk, they must
form after the last accretion outburst.  For this model, the dead zone
is accreted and the disk becomes fully turbulent at a time of around
$10^6\,\rm yr$.  The dead zone does not last the entire disk lifetime
because the active layer surface density is relatively high. As
material drains slowly from the dead zone, the disk soon has
$\Sigma<\Sigma_{\rm crit}$ or $T>T_{\rm crit}$ everywhere in the disk,
and it is fully turbulent. At this time, the amount of material inside
of $1\,\rm AU$ decreases very rapidly. Consequently, with these dead
zone model parameters it is unlikely that there would be enough
material, for sufficient time, for super--Earths to form in the inner
parts of the disk.

\begin{figure}
\begin{center}
\includegraphics[width=8.4cm]{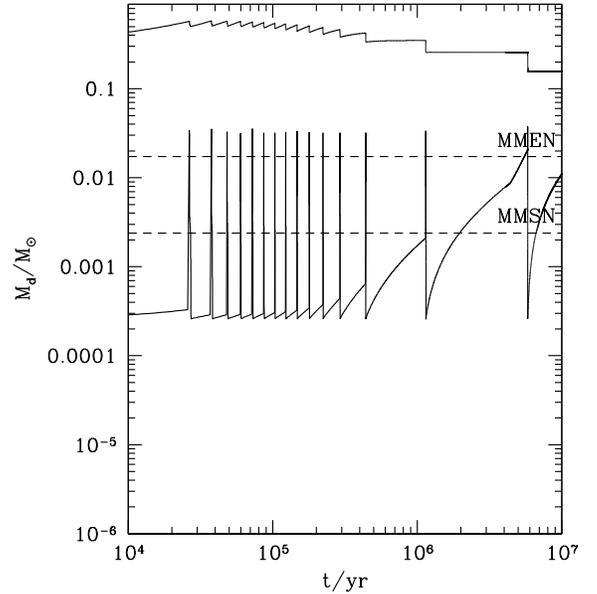}
\end{center}
\caption{Same as Fig.~\ref{fig1} except the disk has a dead zone
  defined by model~1 with $\Sigma_{\rm crit}=20\,\rm g\,cm^{-2}$ and $T_{\rm
    crit}=800\,\rm K$.}
\label{fig3}
\end{figure}

Fig.~\ref{fig3} shows a disk with a dead zone defined by model~1 with
parameters $T_{\rm crit}=800\,\rm K$ and $\Sigma_{\rm crit}=20\,\rm
g\,cm^{-2}$. With the lower active layer surface density, the dead
zone persists for the entire disk lifetime. In the later stages of the
disk lifetime there is sufficient material to form super--Earths {\it
  in situ}. In this model, there is a final accretion outburst at a
time of around $8\,\rm Myr$. However, observationally the lifetime of
protoplanetary disks is only a few Myr
\citep[e.g.][]{Haisch2001,Armitage2003}. Therefore, the disk will
mostly be photoevaporated before this time
\citep[e.g.][]{Clarke2001,Alexander2006,Owen2011}. Photoevaporation of
the outer parts of the disk will cut off the supply of material to the
inner parts. The final accretion outburst shown in the model will most
likely not take place and super--Earths that form while the disk mass
is high will survive.

For this disk model, in which the dead zone lasts the entire lifetime
of the disk, the total mass of the disk is somewhat high, around
$0.1\,\rm M_\odot$ at $t=10\,\rm Myr$. Observations of disk masses are
typically derived from measuring the amount of dust. Assuming a gas to
dust ratio of 100, observed masses are in the range $10^{-3}$ to
$0.1\,\rm M_\odot$ \citep[e.g.][]{williams2014}. Thus, the disk masses
predicted by this model are on the high side. However, in a disk with
a quiescent dead zone, the dust will concentrate to the midplane and
therefore observations may underestimate the mass of the gas disk. The
process of photoevaporation depends upon the accretion flow changing
from being dominated by viscous torques, to being dominated by the
wind mass loss. This transition occurs once the accretion flow through
the disk reaches a critical value which is typically in the range of
$10^{-10}-10^{-8}\,\rm M_\odot\, yr^{-1}$ depending upon the the
dominant photon flux, X--rays, EUV or FUV
\citep[e.g.][]{alexander2014}. In our model, most of the disk mass
(97\%) is in the dead zone at the end of the simulation and so the
accretion flow rate through the disk is small.  While the total disk
mass may be high, the viscous accretion rate is low, and therefore
photoevaporation should be able to efficiently clear the disk.

\begin{figure}
\begin{center}
\includegraphics[width=8.4cm]{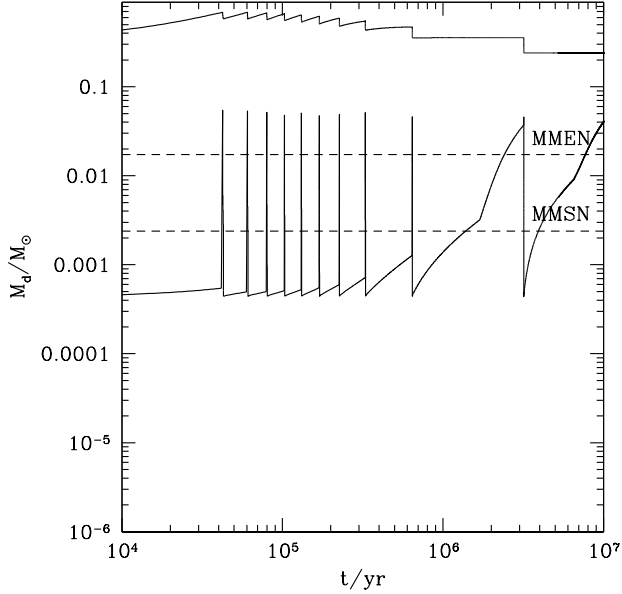}
\end{center}
\caption{Same as Fig.~\ref{fig1} except the disk has a dead zone
  defined by model~1 with $\Sigma_{\rm crit}=20\,\rm g\,cm^{-2}$ and $T_{\rm
    crit}=1400\,\rm K$.}
\label{fig4}
\end{figure}

Fig.~\ref{fig4} shows a disk with a dead zone defined by model~1 with
parameters $T_{\rm crit}=1400\,\rm K$ and $\Sigma_{\rm crit}=20\,\rm
g\,cm^{-2}$. The higher critical temperature allows more time for
super--Earth formation at earlier times.  Again, we suggest that the
final outburst shown will not take place because the disk will be
photoevaporated before this time.

\begin{figure}
\begin{center}
\includegraphics[width=8.4cm]{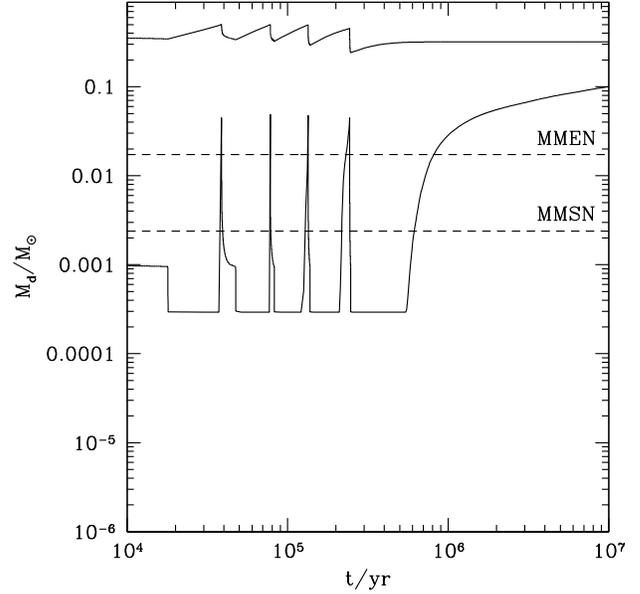}
\end{center}
\caption{Same as Fig.~\ref{fig1} except the disk has a dead zone
  defined by model~2 with$Re_{\rm M,crit}=5\times 10^4$ (bottom
  right).}
\label{fig5}
\end{figure}

Fig.~\ref{fig5} shows a disk with a dead zone determined by
model~2. The critical magnetic Reynolds number is $Re_{\rm
  M,crit}=5\times 10^4$. For this model, the active layer surface
density changes with radius. The inner parts of the disk are almost
entirely non--turbulent and a significant amount of material builds up
there. There is more than enough material for super--Earths to form
{\it in situ}. After a time of about $10^6\,\rm yr$ the disk has more
material than the MMEN in $R<1\,\rm AU$ for the rest of the disk
lifetime.

In our disk model, the inner edge of the dead zone is determined by
the radial distance at which the temperature of the disk drops below
that required for thermal ionisation. It does not change in time
because the temperature profile is not significantly affected by a
dead zone \citep[see for example][]{Lubow2013} unless it becomes self
gravitating.

In conclusion, we have shown that the disk mass inside of $1\,\rm AU$
may be several times that of the MMEN for the later stages of the disk
life and thus formation of super--Earths in this region is possible,
depending on the dead zone parameters.  We should note that we have
described only a representative sample of the results obtained from a
series of simulations.  From all the simulations we find that the dead
zone persists long enough for the formation of super--Earths provided
that $\Sigma_{\rm crit} \lesssim 100\,\rm g\,cm^{-2}$. While the value
of the active layer surface density is still somewhat uncertain, this
value may be representative of protoplanetary disks.  Still, since
different protoplanetary disks may have different values,
super--Earths may form {\it in situ} in some systems but not in
others.

\subsection{Migration of Super--Earths}

Formation of super--Earths farther out in the disk with subsequent
inward migration may be more difficult in a disk with a dead
zone. While the rate of type~I migration is not affected by viscosity
(or the presence of a dead zone), the low viscosity means that the
planet more likely migrates by type~II migration
\citep[e.g.][]{Matsumura2005, Matsumura2006}.  Traditionally it was
thought that there are two gap opening criteria that need to be
satisfied in order for a planet to open a gap in the disk. The first
is the viscous gap opening criterion
\begin{equation}
\frac{M_{\rm p}}{M}  \gtrsim  \left( \frac{40\nu}{R^2\Omega}\right)^\frac{1}{2} \left(\frac{H}{R}\right)^\frac{3}{2}=(40\alpha)^\frac{1}{2}\left(\frac{H}{R}\right)^\frac{5}{2}
\label{mcrit}
\end{equation}
\citep{LP1986}. \cite{Lin1993} suggested that a second criterion, the
thermal criterion, must also be satisfied for a planet to open a
gap. That is, the Hill radius of the planet must be larger than the
disk scale height. However, more recently, it has been found that this
criterion is too strong \cite[e.g.][]{Rafikov2002}. \cite{Zhu2013}
find that due to the non--linear wave steepening, a low mass planet
can open a gap in contradiction to the thermal criterion. In this work
we assume that the gap--opening criterion is determined by the viscous
torque on the disk balancing the tidal torque from the
planet. Rearranging equation~(\ref{mcrit}), we find that a
super--Earth can open a gap in the disk provided that the viscosity
parameter satisfies
\begin{equation}
\alpha \lesssim 1.8\times 10^{-5} \left(\frac{M_{\rm p}}{5\,\rm M_\oplus}\right)^2\left(\frac{M}{1\,\rm M_\odot}\right)^{-2}\left(\frac{H/R}{0.05}\right)^{-5}.
\end{equation}
That is, the dead zone does not have to be entirely turbulence--free
for a super--Earth to open a gap.

%Type~II migration occurs when the planet is large enough to open a
%gap, its mass satisfies \begin{equation} M_{\rm p} \gtrsim 74.9 \,
%\left(\frac{\alpha}{0.01}\right)
%\left(\frac{H/R}{0.05}\right)^{5/2}\left(\frac{M}{1\,\rm
%M_\odot}\right) \, \rm M_{\oplus}.  \end{equation}
Type~II migration typically proceeds on the viscous timescale for the
disk
\begin{equation}
\tau_{\rm \nu}=\frac{R^2}{\nu}=\frac{1}{\alpha (H/R)^2\Omega}
\end{equation}
\citep[e.g.][]{Pringle1981}.
 For typical parameters 
%at the edge of the
%accreting inner region, 
this is given by
\begin{equation}
\tau_{\rm \nu}=\, 7.1 \times 10^{4}\,
\left(\frac{\alpha}{0.01}\right)^{-1} \left(\frac{H/R}{0.05}\right)^2
\left(\frac{M}{1\,\rm M_\odot}\right)^{-\frac{1}{2}}
\left(\frac{R}{5\,\rm AU}\right)^{\frac{3}{2}} \,\rm yr.
\label{tvisc}
\end{equation}
This can be longer than the lifetime of the disk for $\alpha \lesssim
10^{-4}$.  If a super--Earth forms inside a dead zone, it may be
trapped there in the dead region, while migrating only very slowly by
type~II migration.

The disk becomes depleted in time because the infall accretion rate
declines while material continues to accrete on to the star through
the active surface layers. Concomitantly, the dead zone size gradually
decreases. In the end stages of the disk's life, the disk becomes
fully turbulent down to the mid--plane (this occurs when $\Sigma \le
\Sigma_{\rm crit}$ everywhere that $T<T_{\rm crit}$) and the inner
parts of the disk accrete on to the star. When the disk becomes
turbulent, the planet may no longer be able to hold a gap open, and
type~I migration may occur. We discuss this possibility further in the
next Section.

\section{Super-Earths in the Early Solar System}
\label{solarsystem}

The lack of super--Earths in our solar system sets us somewhat apart
from observed exoplanetary systems. There are two possible
explanations for this dearth of super--Earths.  Either conditions in
the solar nebula did not allow for the formation of super--Earths, or
else, they did form but were subsequently somehow removed. Given that
the orbits of the planets in the solar system are coplanar and not
very eccentric, planet--planet scattering does not seem to be a likely
ejection mechanism. Thus, if super--Earths formed, they most likely
fell into the Sun.

If super--Earths form outside of the snow line and migrate inwards
through a gas disk, they affect the composition of the terrestrial
planets, if they migrate slowly enough \citep{Izidoro2014}. Therefore,
if they formed in our solar system, they would have had to migrate
quickly, on a timescale of about $0.01-0.1\,\rm Myr$. If the timescale
is longer than this, then a super--Earth shepherds rocky material
interior to its orbit and depletes the terrestrial planet-forming
zone. Terrestrial planets that form may be volatile rich and are more
likely to be water--worlds that are not very Earth--like.

Consequently, the most probable formation site in our solar system is
in the inner regions, inside of Mercury's orbit. The lack of any
objects in this region may indeed suggest that super--Earths formed
close to our Sun, clearing the region of debris, but subsequently they
fell into the Sun.  The mechanism that pushed the super--Earths into
the Sun could be simply migration through the gas disk. At the end of
the gas disk's lifetime, photoevaporation removes the remaining gas on
a short timescale
\citep[e.g.][]{Clarke2001,Alexander2006,Owen2011}. However, this
process only operates outside of the gravitational radius from the
star, typically around $5-10\,\rm AU$
\citep[e.g.][]{Hollenbach1994}. The inner parts of the disk that we
are interested in accrete on to the central star on a viscous
timescale.
%Alternatively, the super--Earths can begin to migrate while the dead
%zone is still present if the dead zone's inner radius moved to outside
%of the super--Earth's orbit after the super--Earth formation. In
%either case, 
The maximum total surface density present in the disk during this
process is the critical active layer surface density, $\Sigma_{\rm
  crit}$. 
%The amount of material present in the disk determines
%whether there is enough material for the planets to migrate into the
%Sun.
If the critical active layer's surface density is small, then the
planets will not move far during this process. However, a large
critical surface density may be sufficient for a super--Earth to
migrate into the Sun. We therefore find that there is a delicate
balance between the need for a sufficiently large surface density in
the active layer for the planets to migrate into the star, but also a
small enough active layer surface density to allow the planets to form
{\it in situ} in the first place.

For the super--Earth to migrate into the Sun, it must do so on a
timescale shorter than the viscous timescale (the timescale for the
disk to accrete). Equating equations~(\ref{type1}) and~(\ref{tvisc})
we find that the minimum surface density in the disk for the planet to
migrate in to the Sun is
\begin{align}
\Sigma_{\rm min}= \,\,\,& 940.5\,\left(\frac{\alpha}{0.01}\right) \left(\frac{H/R}{0.05}\right)^4 \left(\frac{M}{1\,\rm M_\odot}\right)^{2} \left(\frac{R}{5\,\rm AU}\right)^{-\frac{3}{2}} \nonumber \\
& \times  \left(\frac{a}{1\,\rm AU}\right)^{-\frac{1}{2}}\left(\frac{M_{\rm p}}{5\,\rm M_\oplus}\right)^{-1} \rm g\,cm^{-2} .
\label{sigmin}
\end{align}
Now, if $\Sigma_{\rm crit}>\Sigma_{\rm min}$ then we expect that
super--Earths that form {\it in situ} will migrate into the Sun at the end
of the disk lifetime. On the other hand, if $\Sigma_{\rm
  crit}<\Sigma_{\rm min}$ then there may be some type~I migration, but
not enough to allow the super--Earth to be accreted.  Note that
equation~(\ref{sigmin}) is very sensitive to the disk aspect
ratio. For example, if the disk aspect ratio is decreased by a factor
of two, down to $H/R=0.025$, then we find that $\Sigma_{\rm
  min}\approx 60\,\rm g\,cm^{-2}$.  We speculate that in our
solar system, super--Earths formed in the inner parts of a relatively
cool disk, close to the dead zone inner boundary. There was sufficient
time for them to migrate through the disk to be accreted on to the
Sun. While this outcome is less likely because of the cool
conditions required, it is definitely not impossible and should happen in other
planetary systems also.

We note that an alternative mechanism for pushing the super--Earths
into the Sun is the grand tack \citep{Walsh2011,Batygin2015}. In this
model, Jupiter migrates inwards to $1.5\,\rm AU$ before it gets locked
into resonance with Saturn and then they both move outwards to their
current locations. During this process the innermost super-Earths get
shepherded in to the Sun.
%The formation of Mars with such a small mass
%has caused problems for planet formation theories. When we look at the
%minimum mass solar nebula, we see that Mercury, Mars and the asteroid
%belt have a much smaller mass than would be expected. The probability
%of forming such a small Mars is only around 10\% and this requires a
%non-zero eccentricity for Jupiter and Saturn
%\citep{Fischer2014}. However, this means that it is not impossible. It
%has been suggested that the four terrestrial planets must have formed
%out of a narrow annulus of rocky debris in the orbital range
%$0.7-1\,\rm AU$ \citep{Hansen2009}. This led to the suggestion of the
%outer truncation being caused by Jupiter migrating inwards to
%$1.5\,\rm AU$ before reversing its direction of migration while being
%locked into resonance with Saturn
%\citep{Walsh2011}. \cite{Batygin2015} further suggested that early
%super--Earth planets that formed in the inner solar system could have
%been forced into the Sun during the grand tack. This explains the
%lack of anything inside of Mercury's orbit. 
Our disk model which includes a dead zone also provides an alternative
explanation for the cleared inner regions of our solar system and the
lack of super--Earths. Furthermore, the small masses of Mercury and
Mars can be explained if the terrestrial planets form from a narrow
annulus of rocky debris in the orbital range $0.7-1\,\rm AU$
\citep{Hansen2009}. Our model can explain the inner truncation radius
for this annulus as being where the super--Earths cleared the
material.

%But, is the grand tack necessary? The survival of super--Earths in the
%inner parts of the disk may be dependent on the time of their
%formation. If they form while there is a large amount of turbulent
%material in the gas disk then they could migrate into the Sun on a
%short timescale. However, if they form in the late stages of the disk
%lifetime then they can survive.....

%Icy region in inner part of disk...

\section{Discussion}
\label{discussion}

The MMEN is criticized because of the unusually high amount of
material required for the {\it in situ} formation of
super--Earths. However, sub-mm observations of disks measure the
properties of the outer disk while the inner parts of the disk are not
very well constrained. Furthermore, \cite{Ogihara2015} suggest that
{\it in situ} formation cannot operate unless type~I migration is
suppressed in the region inside of $1\,\rm AU$. A disk model with a
dead zone (such as the one proposed in this work) not only provides
sufficient material, but it also suppresses the rate of migration in
that region.

\cite{Raymond2014} argue that super--Earths could not have formed {\it
  in situ}. They use observations of systems that contain three or
more planets to construct an MMEN and find that the surface density
profile $\Sigma \propto R^\sigma$ has $-3.2<\sigma<0.5$. They suggest
that because viscous accretion disk models have difficulty in
reproducing such extreme profiles that the super--Earths could not
have predominantly formed {\it in situ}. These conclusions are based on
a steady state disk model with the temperature profile dominated by
stellar irradiation \cite[e.g.][]{Chiang1997}.  However, we suggest
that the extreme values could be a result of a discontinuous surface
density distribution brought about by the presence of a dead zone. In
the dead zone material builds up and the surface density in the
innermost MRI active parts of the disk may be significantly lower than
that farther out in the dead zone. If this is the case, then one
cannot smooth the profile from three planets out to a continuous
surface density distribution. Consequently, the most important factor
is the amount of material in the inner regions available for planet
formation, rather than its distribution.
% In this work we have
%concentrated on the amount of material inside of $1\,\rm AU$.

Because planets that form {\it in situ} are expected to require an
unusually large amount of solids in the inner parts of the disk, the
prediction is that the metallicity of the host star must be higher in
such cases. \cite{Zhu2015} found that close-in super-Earths are more
likely to be found around metal-rich stars. Similarly, the widely
separated super-Earths are more often around metal--poor stars. As
metal-rich stars have more solid material closer to the star within
the disk, this appears to favour the {\it in situ} formation. However,
there may be be two populations of super--Earths because there is a
distinct transition in orbital period observed, rather than a smooth
transition.

\cite{Chatterjee2014} proposed a mechanism to form the close in
super--Earths at the pressure trap at the inner edge of a dead zone,
known as inside--out planet formation. Once a planet forms there, the
inner edge of the dead zone moves out allowing planet formation
further out \citep{Chatterjee2015,Hu2015}. This model requires a high
rate of supply of pebbles to the inner disk.  We suggest that a dead
zone model can provide all the required material for planet formation
without the need for accumulation. However, the inside--out planet
formation mechanism potentially increases the range of possible dead
zone parameters that are able to form the super--Earths {\it in situ}.

In this work we have considered only the formation of super--Earth
planets during the gas disk lifetime. It is entirely possible that
some super--Earth planets may form after the gas disk has dissipated
and we expect these to have a high density, similar to the terrestrial
planets in our solar system. However, this is not the only mechanism
to form high density super--Earths.  A super--Earth that forms in a
dead zone with a small active layer will not accrete much material
once it has carved out a gap. The only accretion on to the planet is
through the active layer. Thus, planets that form within a dead zone
are likely to be less gaseous and more dense than planets that form
farther out in a fully turbulent disk and migrate inwards.  This may
explain the weak dependence of density on semi--major axis observed in
Fig.~1.  However, the gaseous atmosphere of a close--in super--Earth
could be stripped from the planet, by tidal evolution or evaporation,
leaving only the solid core behind
\citep[e.g.][]{Schaefer2009,Jackson2010}. Consequently, as we cannot
tell the difference between these two mechanisms in the observed
exoplanets, it is difficult to make firm conclusions as to which
super--Earths form by which mechanism. Nevertheless, the existence of
these different mechanisms may explain the range in the densities.

\section{Conclusions}
\label{conc}

There are two possible formation locations for observed close--in
super--Earths in exoplanetary systems: either {\it in situ}, or
farther away from the star followed by migration to their observed
location.  We find that a disk that contains a dead zone (a region of
low turbulence) may have sufficient material for the planets to form
{\it in situ} although it depends upon the dead zone parameters. In
order for the dead zone to last long enough for super--Earths to form,
the active layer surface density must be sufficiently low,
$\Sigma_{\rm crit}\lesssim 100\,\rm g\,cm^{-2}$. If the active layer
surface density is too large, the disk becomes fully turbulent before
there is sufficient time to form the super--Earths. Migration of
super--Earths through a dead zone is very slow and thus formation
farther out in a disk with a dead zone is more difficult.

We find that a fully turbulent protoplanetary disk model does not have
sufficient material in the inner parts of the disk for {\it in situ}
formation of super--Earths. In this case, the only possible formation
mechanism involves migration from farther out in the disk. The fast
rate of migration in a fully turbulent disk lends itself to this
scenario. We suggest that the observed large range in super--Earth
compositions may be the result of these two very different formation
locations.

The lack of super--Earths in our solar system is somewhat puzzling
given that more than half of observed exoplanetary systems contain
one. However, the fact that there is nothing inside of Mercury's orbit
may not be a coincidence. {\it In situ} formation of super--Earths in
that region could have cleared the solid material. The super--Earths
would have had to subsequently fall into the Sun.  This is possible if
the active layer surface density is sufficiently large that during the
final accretion process there was enough material in the disk for the
planets to migrate into the Sun. In order to satisfy both this
constraint, and the constraint that the dead zone must last throughout
the disk lifetime, requires a sufficiently cool disk during the final
accretion process. The level of fine--tuning required is certainly
possible, but we don't expect it to happen in all systems and this
can explain why the solar system is somewhat special in its lack of
super--Earths.

\section*{Acknowledgments} 

We thank an anonymous referee for useful comments. This research has
made use of the Exoplanet Orbit Database and the Exoplanet Data
Explorer at exoplanets.org.

\bibliographystyle{apj}

\label{lastpage}
\end{document}